# Energy-gap modulation of boron nitride nanoribbons by transverse electric fields: First-principles calculations

Zhuhua Zhang and Wanlin Guo[*]

*Institute of Nano Science, Nanjing University of Aeronautics and Astronautics, Nanjing 210016, People's Republic of China*

Systematic *ab initio* calculations show that the energy gap of boron nitride (BN) nanoribbons (BNNRs) with zigzag or armchair edges can be significantly reduced by a transverse electric field and completely closed at a critical field which decreases with increasing ribbon width. In addition, a distinct gap modulation in the ribbons with zigzag edges is presented when a reversed electric field is applied. In a weak field, the gap reduction of the BNNRs with zigzag edges originates from the field-induced energy level shifts of the spatially separated edge-states, while the gap reduction of the BNNRs with armchair edges arises from the Stark effect. As the field gets stronger, the energy gaps of both types of the BNNRs gradually close due to the field-induced motion of nearly free electron states. Without the applied fields, the energy gap modulation by varying ribbon width is rather limited.

# I. INTRODUCTION

As a quasi-two-dimensional insulator, hexagonal boron nitride (*h*-BN) has been widely used in electronic devices due to its high thermal stability.[1] Recently, the single-layered *h*-BN atomic crystals were discovered experimentally,[2] which had been shown to possess excellent electronic and optical properties.[3,4] From the two-dimensional crystal, BN nanoribbons (BNNRs) with different widths and selected edge directions can be produced experimentally in the same way as graphene nanoribbons.[5] The graphene nanoribbons, especailly with zigzag edges, have been found to be conductive,[6] magnetic[7] and even half-metallic[8] materials, mainly due to the localized edge-states originated from the $\pi$ orbitals of the atoms near the edges[9]. However, as an analogy to graphene ribbons structurally, the BNNRs are much less studied. A theoretical study has found that the band structures of the BNNRs with zigzag edges (Z-BNNR) exhibits flat-band edge-states and the energy gap monotonically decreases with increasing ribbon width,[10] but systematic study on the electronic structures of BNNRs is still lacking. On the other hand, the electric field has been proved to be able to modulate the electronic structures of many nanoscale materials.[11,12] Recent calculations have shown that BN nanotubes undergo an insulator-metal transition under a transverse electric field, which depends weakly on the chirality but strongly on the nanotube diameters.[13,14] However, the BNNRs with different shaped edges should have different responses to the external electric field. Particularly in the Z-BNNR, the break of the lattice symmetry induces a spontaneous transverse electric polarization due to the existence of the asymmetrical ionic potential,[15] which makes the Z-BNNRs obviously different from the BN nanotubes. Thus, the modulation of transverse electric fields on the electronic properties of the BNNRs also needs to be clarified.

In this work, we show through *ab initio* density-functional theory calculations that the energy gap of BNNRs changes slightly with varying width, but can be significantly reduced by a transverse external electric field ($E_{\text{ext}}$). It is also shown that an insulator-metal transition can be realized at a critical field strength which decreases remarkably with increasing ribbon width. Therefore, the electric field-induced energy gap modulation should have potential applications.

## II. MODEL AND METHODOLOGY

All the calculations are carried out using ultrasoft pseudo-potentials with a plane-wave basis,[16] which is restricted by a cutoff energy of 434.8 eV. In the calculations, the local density approximation (LDA) is employed to describe the exchange-correction potential. A one-dimensional periodic boundary condition is applied along the ribbon edge, and one unit cell of the nanoribbon with various width is included. Both of the distances between edges and between planes of two adjacent ribbons are at least 1.1 nm, which is large enough to eliminate the effect of direct ribbon-ribbon interaction. The Brillouin-zone integration is sampled by up to 10 special *k* points for atomic structure relaxation and total 25 *k* points for electronic structure calculation. A periodic saw-tooth type potential perpendicular to the ribbon edge is used to simulate the transverse electric fields in a supercell, and such a potential is homogeneous along the ribbon edges[17]. The conjugate gradient method is used to optimize the geometry, and all the atoms in the unit cell are fully relaxed until the force on each atom is less than 0.1 eV/nm.

### III. ENERGY GAP MODULATION BY RIBBON WIDTH

The Z-BNNRs are classified by the number of zigzag chains ($n_z$) across the ribbon width and denoted as Z-$n_z$ [Fig. 1(a)], while the BNNRs with armchair edges (A-BNNRs) are classified by the number of dimer lines ($n_a$) across the ribbon width and denoted as A-$n_a$ [Fig. 1(b)]. Our calculations show that the energy gap of the Z-BNNRs is indirect and decreases with increasing width as shown in Fig. 1(c); while the energy gap of the A-BNNRs is direct and the gap variations versus width exhibit three distinct family behaviors[Fig. 1(d)], for ribbons with $n_a= 3p$ and $n_a=3p+1$ (where $p$ is a positive integer) the energy gap decreases with increasing $n_a$, whereas for ribbons with $n_a=3p+2$, the energy gap increases monotonically with increasing $n_a$.

#### A. Zigzag BN nanoribbon

The monotonical decrease in the energy gap of Z-$n_z$ ribbons should be related with the two edge-states localized at the B and N edges, respectively,[10] and can not be explained by quantum confinement effects because the energy gap is smaller than that of a BN sheet when $n_z>5$. With increasing $n_z$, the site energy at the B (N) edge is lowered (raised), and thus more charges are depleted (accumulated) at the B (N) edge.[10] The charge transfer from the B edge to the N edge is thus enhanced by increasing the ribbon width, which should be the origin of the monotonical reduction in the energy gap. Taking the Z-08 ribbon as an example, its band structure is shown in Fig.2 (a). The highest occupied molecular orbital (HOMO) is located at $k_H$ while the lowest unoccupied molecular orbital (LUMO) lies at X. As shown in Fig. 2(b), the HOMO consisting of $\pi$ states corresponds to the edge-state decaying inward from the N atoms at N edge while the LUMO consisting of $\pi^*$ states corresponds to the edge-state localized on the B atoms at B edge. As the ribbon width increases, the enhanced charge transfer from the B edge to the N edge makes the HOMO (LUMO) shift upward (downward), resulting in decreasing energy gap.

### B. Armchair BN nanoribbon

Next, to clearly understand the width-dependent variations in the energy gap of A-BNNRs, we employed a tight binding (TB) model with first- and second-nearest-neighbor interactions [18] and transformed the A-BNNR to a topologically equivalent lattice model as described in Ref.19. The TB calculations show that the energy gap of the A-$n_a$ ribbons with $n_a=3p+2$ nearly keeps the same as that of a BN sheet regardless of its width while the energy gaps of the ribbons with $n_a=3p$ and $n_a=3p+1$ monotonically decrease with increasing $n_a$. All the gaps in the three families would approach the gap value of the infinite BN sheet as the ribbon width increases. The results suggest that the quantum confinement is an important factor in the variation of the band gap. The three family behaviors of the gap variation are due to the quantized wave vector along the width direction. Similar phenomenon has been found in previous researches on the zigzag carbon nanotubes and graphene ribbons with armchair edges.[19,20] However, the TB calculations also show that the energy gaps of the A-09, A-10 and A-11 ribbons are 4.541 eV, 4.504 eV and 4.492 eV, respectively, with a hierarchy of the gap size given by A-09>A-10>A-11, inconsistent with those of 4.514 eV, 4.528 eV and 4.466 eV (A-10>A-09>A-11) obtained by the first-principles calculations. This discrepancy is originated from that the edge effects are not considered in the TB calculations. Because the edge atoms of the ribbon are terminated by hydrogen atoms, the bond lengths near the two edges are shorter than those in the middle of the ribbon. When such edge effects are taken into account by modifying the transfer integrals between atoms near the edges,[21] we find that the energy gaps of the A-09, A-10 and A-11 ribbons are 4.517 eV, 4.530 eV and 4.462 eV, respectively, with a hierarchy of the gap size given by A-10>A-09>A-11. Both the gap values and hierarchy are in good agreement with the first-principles calculations. This analysis proves the importance of the edge effects for the width-dependent gap variations of the A-BNNRs, and also provides the physical origin of why the energy gap of the ribbons with $n_a=3p+2$ is lower than that of the BN sheet.

**IV. ENRRGY GAP MODULATION BY TRANSVERSE ELECTRIC FIELD**

According to the above discussions, the BNNR is an insulator regardless of its width and edge shape because of the large energy gap, though the gap variations versus width are diversified. To further obtain a significant modulation in the band gap of the BNNRs, an external electric filed is transversely applied. We define the positive direction of the applied electric field to be from the B edge to the N edge as depicted in Fig. 1(a).

### A. Zigzag BN nanoribbon

When a positive electric field is applied, the energy gap of the Z-08 ribbon decreases with increasing $E_{ext}$ and is eventually closed at a field around 6.5 V/nm as shown in Fig. 2. As discussed above, there are two edge-states localized at the two ribbon edges, which exponentially decay into the center of the ribbon, similar to those found in graphene nanoribbons.[9] The decay rate can be measured by the distance of $k_X$ - $k_H$ [Fig. 2(a)]. It is shown that $k_X$ - $k_H$ is shortened by the applied electric field and the location of the HOMO gradually shifts to X, while the LUMO is always localized at X. The reason is that the electrostatic potential is raised at the N edge and lowered at the B edge as $E_{ext}$ increases, leading to more charge accumulated at the N edge but more depleted from the B edge as shown in Fig. 2(b). This redistribution of the edge-states causes the gap reduction. However, when $E_{ext}$ is above 4.1 V/nm, the gap reduction is maintained by the rapid drop of the NFE band as shown in Fig. 2(a). In the transverse electric field, the NFE states also move to the left side of the unit cell where the electrostatic potential is lowered [Fig. 2(b)]. Since the NFE states reside far from the ribbon plane, they are weakly bound to the plane of the BNNR,[7,22] unlike with the $\pi$ states which are locked to the plane of the ribbon. Thus, the charge density of the NFE states is easily redistributed under the transverse electric field, making the NFE band shift downward more rapidly than the LUMO at X (LUMO$_X$). When $E_{ext}$ reaches up to 6.5 V/nm, the NFE states have been completely accumulated to the left side of the ribbon, giving rise to the band gap closure. Here, it should be noted that the unit cell dimensions used in our calculations are sufficiently large so that the effect of finite space on the positions of the NFE band at different values of $E_{ext}$ can be negligible.

Under a reversed electric field, a distinct energy gap modulation is presented for the Z-08 ribbon as shown in Fig. 3. The energy gap increases firstly when the field strength is less than 0.5 V/nm and then rapidly decreases thereafter. In this case, once the applied field is turned on, the electrostatic potential is raised at the B edge and lowered at the N edge. Correspondingly, the LUMO$_X$ (HOMO) shifts upward (downward) for a small energy distance. In this case, the edge-states can not be shifted away immediately under the weak $E_{ext}$ due to the existence of the spontaneous transverse electric polarization[15], which is against the direction of the electric field. Consequently, the $k_X$-$k_H$ is elongated a little, and the density of the edge-states slightly decreases as shown in Fig. 3(b). As the field gets stronger, the LUMO$_X$ gradually shifts to the leftmost B atoms of the unit cell while the HOMO shifts to the rightmost N atoms, following a slow decrease in the energy interval between the HOMO and the LUMO$_X$. During this course, the $k_X$-$k_H$ is slightly shortened as shown in Fig. 3(a) because the external field strength is large enough to overcome the intrinsic transverse electric polarization and thus redistributes the edge-states. However, the accumulation of the NFE states induced by the electric field to the left side of the unit cell leads to a rapid drop of the NFE band [Fig. 3(a)], which promotes the gap reduction and closes it at a field around 5.5 V/nm. This is similar to the discussion on the positive electric field. Furthermore, it is indicated that the degree of the drop of the NFE band under the same field strength is more significant for the reversed electric field than for the positive field.

In Fig. 3(a), we also find a direct gap at $\Gamma$ under field strength of 6 V/nm. The direct gap is gradually formed between the NFE band and a guided band which rises with increasing $E_{\text{ext}}$ [see Fig. 3(a)]. According to the calculation of the Bloch wave functions, it is found that the states related with the guided band are mainly from the edge B-H $\sigma$ bonds as shown in the inset of Fig. 3(a), and can be regarded as the effect of H $1s$ states on the band structure of the Z-BNNR. More charge density of the states is distributed on H atoms due to its larger electron affinity. Under the reversed electric field, the site energy at the B-H bonds is pushed up, leading to the guided band shift close to the Fermi level. The resultant direct band gap in the Z-BNNRs should be beneficial for optoelectronic and photonic applications.

### B. Armchair BN nanoribbon

We now discuss the effect of $E_{\text{ext}}$ on the band structures of the A-BNNRs. The band structures of the A-17 BNNR under different positive $E_{\text{ext}}$ are calculated as shown in Fig. 4(a). It is shown that the band gap is direct at $\Gamma$, and also can be reduced and closed by $E_{\text{ext}}$. In the band structure of the A-17 ribbon, the HOMO consisting of $\pi$ states is uniformly distributed on the N atoms while the LUMO consisting of $\pi^*$ states is resided on the B atoms. Under a weak $E_{ext}$, the energy levels shift and split, leading to a reduction in the band gap. This effect is known as the Stark effect, which has been observed in previous studies on carbon and BN nanotubes.[12, 13] As the electric field increases, the HOMO shifts to the right side of the ribbon where the electrostatic potential is raised, while the LUMO shifts to the left side where the electrostatic potential is lowered as shown in Fig. 4 (b), which further reduces the band gap. On the other hand, as discussed above, the NFE states are more easily redistributed than the $\pi^*$ states due to its high sensitivity to $E_{ext}$, which induces the rapid drop of the NFE band in energy relative to other conduction bands as shown in Fig 4. In consequence, the energy gap of the A-17 ribbon is closed at $E_{ext}$=5 V/nm. However, in this type of ribbons, no distinct difference is found when the electric field direction is reversed.

### C. Ribbon width dependence of the gap modulation

We also find that the gap modulation in BNNRs induced by the transverse electric field is width-dependent. In Fig. 5, it is shown that the wider the ribbon is, the more rapidly the gap decreases. This width dependence can be explained by the electrostatic potential difference between the two ribbon edges, which is proportional to the size of the BNNR. Due to the computational cost, the width range in the figures are limited, and the critical electric field to close the band gap of the BNNRs in this width range is high and seems to be difficult to reach in laboratory.

Nevertheless, with further increasing ribbon width, the critical field strength will decrease rapidly. For example, band gap closure in the Z-26 ribbon with a width of 5.02 nm can be realized at $E_{ext}$=1 V/nm, which is practically available. In experiments, the ribbon widths usually reach up to tens or even hundreds of nanometers, which would make the band gap sensitively modulated and easily closed by the applied electric fields.

The asymmetrical band gap modulation of the Z-BNNRs is also presented clearly in the top panel of Fig. 5 if one compares the curves with positive and reversed electric fields. It is shown that the gap reduction in the Z-08 BNNR under a positive field is faster (slower) than that under a reversed field when $E_{ext}$ is below (above) 1.1 V/nm. Therefore, 1.1 V/nm can be regarded as a critical electric field to select a way for the rapidest gap modulation of the Z-08 ribbon. This critical field would increase with increasing ribbon width, and hence would have practical importance in the gap modulation of Z-BNNRs with larger size.

## V. CONCLUSION

In summary, our first-principles calculations show that the BNNR is an insulator regardless of its width and edge shape, however, an insulator-metal transition in the ribbon can be gradually realized through an increasing transversely applied electric field. It is important that the critical electric field for the transition can decrease to a practical range with increasing ribbon width. A reversed electric field can induce a distinct band gap modulation in the Z-BNNRs. When a weak field is applied, the gap reduction of the Z-BNNRs is due to the field-induced redistribution of edge-states, while the energy gap narrowing in the A-BNNRs arises from a Stark effect. As the field strength increases, the redistribution of the NFE states induced by electric field makes the energy gap gradually closed for both types of the BNNRs.

We are aware that the density-functional theory usually underestimates the band gap of semiconductors and insulators.[23] Nevertheless, the GW correction is expected to just increase the threshold field for band gap closure in the BNNRs and should not change the presented trends of the energy gap modulation, because the field-induced modulation on the band gap is induced by the redistribution of either $\pi$ states or the NFE states. The charge redistribution should not be qualitatively affected by the GW self-energy correction.[24] Therefore, the convenient gap modulation in BNNRs is promising for applications in future nanodevices.

## ACKNOWLEDGEMENTS

This work is supported by 973 Program (2007CB936204), the Ministry of Education (No. 705021, IRT0534), National NSF (10732040), Jiangsu Province NSF of China and Jiangsu Province Scientific Research Innovation Project for Graduate Student (CX07B_064z).

FIG. 1. (Color online) Diagram of (a) Z-$n_z$ and (b) A-$n_a$ BNNRs, with widths of $n_z = 8$ and $n_a = 15$, respectively. In these BN structures, all edges are passivated with hydrogen atoms. The longitudinal direction is the infinitely extended direction. The variation of band gaps of Z-$n_z$ and A-$n_a$ BNNRs is drawn as a function of widths in (c) and (d), respectively. Blue dot lines denote the energy gap of an isolated BN sheet along the corresponding directions. $p$ is a positive integer in (d). The positive direction of $E_{ext}$ is denoted by a big arrow in (a) and (b) for subsequent discussions.

FIG. 2. (Color online) Effects of positive electric field on the electronic structures of the Z-BNNR. (a) Band structures of Z-08 BNNR at $E_{ext} = 0$, 4 and 6.5 V/nm. The HOMO is located at $k_H$. In all figures, the Fermi level is set to zero and the NFE band is guided by a red dash dot line. (b) Charge densities of HOMO and LUMO at $E_{ext} = 0$, 3 and 6 V/nm. The relative magnitude of the charge density is indicated in the right bar.

FIG. 3. (Color online) Variation of band structures of Z-08 BNNR under reversed electric field. (a) The band structures of Z-08 BNNR at $E_{ext} = -0.5$, -4 and -6 V/nm. The NFE and the B-H $\sigma$ band are indicated by red dash dot and green solid lines, respectively. Isosurface plot for the B-H $\sigma$ state is shown in the inset. (b) Charge

densities for HOMO and LUMO at X at $E_{ext}$ = -0.5, -4 and -6 V/nm. All other stipulations follow the convention described in Fig. 2.

FIG. 4. (Color online) Effects of positive electric field on the electronic structures of A-BNNR. (a) Band structures of A-17 BNNR at $E_{ext}$ = 0, 3 and 5 V/nm. (b) Charge densities for HOMO and LUMO at $E_{ext}$ = 0, 2.9 and 5 V/nm. All other stipulations follow the convention described in Fig. 2.

FIG. 5. (Color online) Dependence of the gap reduction on system size. Top and bottom panels present the cases of Z-BNNRs and A-BNNRs, respectively. Dot lines in the top panel indicate the cases under the reversed electric fields.